\documentclass[a4paper, 10 pt, conference]{ieeeconf}  

\IEEEoverridecommandlockouts                              
\usepackage{amsmath} 
\usepackage{amssymb}  
\usepackage{bbm}
\usepackage{graphicx}

\usepackage{array}
\newcolumntype{C}[1]{>{\centering\arraybackslash}p{#1}}

\usepackage[font=small,labelfont=bf]{caption}

\title{\LARGE \bf
Gaussian Processes for HRF estimation for BOLD fMRI
}

\author{Michael Eickenberg, Aina Frau-Pascual, Andr\'es Hoyos-Idrobo
}
\graphicspath{{./images/}}

\begin{document}

\maketitle
\thispagestyle{empty}
\pagestyle{empty}

\begin{abstract}

We present a non-parametric joint estimation method for fMRI task activation values and the hemodynamic response function (HRF). The HRF is modeled as a Gaussian process, making continuous evaluation possible for jittered paradigms and providing a variance estimate at each point. 

\end{abstract}

\section{INTRODUCTION}


The hemodynamic response function (HRF) in Blood-oxygen-level-dependent~(BOLD) functional MRI~(fMRI) is the impulse response to neural activity. Under a linear time invariant system hypothesis, the BOLD response is a convolution of temporal neural dynamics and the HRF. Its characteristic shape can be described in a stylized manner as a difference of gamma functions. However, in reality, not all HRFs adhere to this shape and vary significantly across brain regions and subjects \cite{Handwerker04,Badillo13}. Taking into account these variations can lead to finer predictive models \cite{Pedregosa15}.

Several works have focused on the estimation of the HRF using regularized finite impulse response (FIR) \cite{Goutte00}, constrained linear basis sets \cite{Woolrich04} such as the Fourier basis \cite{Kay08} or splines \cite{Zhang14}. The assumption of spatially locally constant HRFs has been used in \cite{Makni05,Vincent10} in a Bayesian framework to counter the intrinsically low SNR. 

When event presentation times and measurement times can be placed on regular temporal grids, the HRF can be estimated as a signal vector sampled at the greatest common divisor of the two grid steps.
Severe oversampling may be necessary in order to accommodate both grids. This can be dealt with using a smoothness prior on the hrf vector, as in~\cite{Ciuciu03}.

A solution to both the oversampling issue and the irregular sampling setting is to assume a continuous HRF. This can be realized e.g. by fitting a finite linear combination of pre-defined continuous basis functions. Using a probability distribution across functions, e.g. a Gaussian process prior with a smoothness-inducing covariance kernel~\cite{Rasmussen04}, provides access to the full function space and allows variance estimation to quantify uncertainty.

\paragraph*{Contributions} in this work, we cast the problem of estimating the HRF as Gaussian process~(GP). Additionally, we propose an alternating optimization method to find a suitable solution, which iterates between the detection of neural activation and HRF estimation. 

\section{The model and the method}
For the measured signal we assume a classic general linear model (GLM), in which we additionally suppose the hemodynamic response function to be the realization of a Gaussian process:
$$h\sim\mathcal{GP}(\mu, k),$$
where $\mu$ is a mean function and $k$ is a covariance kernel. We assume that it is defined on the interval $[0, L]$ and extended by 0 outside this interval if necessary (i.e. $h = h\mathbbm{1}_{[0, L]}$).
\paragraph*{The signal model}
We assume an fMRI experiment with $P$ different conditions. Each condition $1\leq p\leq P$ is presented a certain number $M_p$ times, each of which we call an event. For this work we consider event presentation to be instantaneous, i.e. of duration 0, but generalization to non-zero duration is possible. The \textit{event indicator functions},
$$e_p(t) = \sum_{m=1}^{M_p}\alpha_{m, p}\delta(t - \tau_{m, p}),$$
encode when which event takes place and with which intensity: $\tau_{m, p}$ is the time of the $m^\textrm{th}$ event of condition $p$, and $\alpha_{m, p}$ is a modulation value that can indicate event intensity (set to 1 by default). 
We call \textit{continuous regressors} the functions
$$x^h_p(t) = e_p\ast h(t).$$
These continuous regressors indicate the BOLD activity due to condition $p$ when modeling the BOLD response as a linear time invariant system with impulse response $h$. 
Given a set of activation values $\beta_p$ for each condition and adding Gaussian i.i.d  white noise of variance $\sigma^2$, $\varepsilon(t)\sim\mathcal N(0,\sigma)$, the continuous signal model for one voxel can be written as 
$$y(t) = \sum_{p=1}^P\beta_px^h_p(t) + \varepsilon(t).$$
In fMRI we have a discrete number of measurements at timepoints $t_n, n=1,\dots, N$, which usually live on an evenly spaced grid: $t_n = nTR$, where $TR$ is the repetition time (e.g. 2s). Evaluating the continuous signal model in these measurement points gives us the discrete signal model
$$y_n = y(t_n) = \sum_{p=1}^P\beta_px^h_p(t_n) + \varepsilon(t_n) = \sum_{p=1}^P\beta_px^h_{n,p} + \varepsilon_n,$$
which can be summarized in vectorial notation as
$$y = X_h\beta + \varepsilon,$$
with $X_h = (x^h_{n, p})_{n, p}\in\mathbb{R}^{N\times P}$ and $\varepsilon = (\varepsilon_n)_n\in\mathbb R^N$.

This is a noisy bilinear signal model, as it is linear in $\beta$ and linear in $h$. At $h$ fixed, the optimal $\beta$ is the least squares solution $X_h^+y$, where $X_h^+$ is the pseudoinverse of $X_h$. At $\beta$ fixed, one observes that the $y_n$ are noisy linear combinations of function evaluations of the Gaussian process $h$. This lends itself to an alternating optimization scheme. Before elaborating it, we make a brief detour into the estimation of a Gaussian process given linear combinations of function evaluations.
\subsection{Linear combinations of Gaussian process evaluations}\label{lincombGP}
The goal of this section is to estimate conditional mean and covariance of a Gaussian process given noisy linear combinations of function evaluations.

Consider $f\sim\mathcal{GP}(\mu, k)$. A multiple measurement $\varphi_n$ can be written as
$$\varphi_n = \sum_{i\in I_n}\eta_{i, n}f(x_{i,n}) + \varepsilon_n,$$
where the $I_n$ are index sets which can be of varying size. Conditioning on the outcomes of these measurements leads to a conditional Gaussian distribution
with a conditional mean and covariance, which can be evaluated in new points $x_k'$ or even new linear combinations thereof:
$$(f(x_k'))_k\big |\{\varphi_n = a_n\}_n\sim\mathcal N(\mu^\textrm{cond}(x'), \Sigma^\textrm{cond}(x')).$$
In other words, and according to the definition of Gaussian processes, any finite number of evaluations follows a Gaussian distribution. Let $z = (z_1^T, z_2^T)^T\in\mathbb R^{N + N'}$, where $z_1 = (\varphi_1,\dots, \varphi_N)^T$ the noisy linear combinations of measurements with known outcomes and $z_2 = (f(x_1'),\dots,f(x_{N'}'))$ a set of function values (or possibly linear combinations thereof, omitted here for simplicity) with unknown outcomes. Then the joint distribution is $z\sim\mathcal N(\mathbb E(z), \textrm{cov}(z))$. We have $\mathbb E(\varphi_n) = \sum_{i\in I_n}\eta_{i, n}\mu(x_{i,n})$. Further, we set
\begin{eqnarray*}
\Sigma^{11}_{mn} & = & \textrm{cov}(\varphi_n, \varphi_m)\\ {} & = & \textrm{cov}(\sum_{i\in I_n}\eta_{i, n}f(x_{i,n}) +\varepsilon_n, \sum_{j\in I_m}\eta_{j, m}f(x_{j,m})+\varepsilon_m) \\ {} & = &  \sum_{i\in I_n, j\in I_m}\eta_{i, n}\eta_{j, m}\textrm{cov}(f(x_{i, n}), f(x_{j, m})) + \sigma^2\delta_{mn}\\ {} & = & \sum_{i\in I_n, j\in I_m}\eta_{i, n}\eta_{j, m}k(x_{i, n}, x_{j, m}) + \sigma^2\delta_{mn}.
\end{eqnarray*}
Similarly, $\Sigma^{21}_{kn}=\textrm{cov}(f(x_k'), \varphi_n) = \sum_{i\in I_n}\eta_{i, n}k(x_k', x_{i, n})$, and $\Sigma^{22}_{kl}\textrm{cov}(f(x_k'), f(x_l')) = k(x_k', x_l')$. We can then say that $\textrm{cov}(z) = \Sigma$, with
$$\Sigma=\left(\begin{array}{c|c}\Sigma^{11} & \Sigma^{21T}\\\hline\Sigma^{21} & \Sigma^{22}\end{array}\right).$$
Conditioning on the measurement values $\{\varphi_n = a_n\}$ yields the conditional Gaussian distribution
$$\mu^\textrm{cond} = \mathbb E[z_2] + \Sigma^{21}{\Sigma^{11}}^{-1}(a - \mathbb E[z_1]),$$
$$\Sigma^\textrm{cond} = \Sigma^{22} - \Sigma^{21}{\Sigma^{11}}^{-1}{\Sigma^{21}}^T$$
for $z_2$.
\subsection{A two-step algorithm for HRF estimation using Gaussian processes}
We maximize the conditional loglikelihood
$$\log p(y|h, \beta) = -\frac{n}{2}\log(2\pi\sigma^2) - \frac{\|y - X_h\beta\|^2}{2\sigma^2}$$
with respect to $h$ and $\beta$ alternatingly. The optimization in $\beta$ at $h$ fixed is obtained by solving the least squares problem as $\hat\beta = X_h^+y$. The optimization in $h$ at $\beta$ fixed amounts to conditioning the Gaussian process on the noisy linear combinations of measurements.

Let $\rho_{m, p}^n = t_n - \tau_{m, p}$. All $\rho_{m,p}^n$ which lie in $[0, L]$ are time points in which $h$ is evaluated in order to construct the GLM. We would thus like to obtain an estimate of these values. However, the only measurements we have access to are
$$y_n = \sum_{p=1}^P\sum_{m=1}^{M_p}\alpha_{m, p}\beta_p\mathbbm{1}_{[0, L]}(\rho_{m,p}^n)h(\rho_{m,p}^n) + \varepsilon_n.$$
By setting $\eta_{m, p}^n = \alpha_{m,p}\beta\mathbbm 1_{[0, L]}(\rho_{m,p}^n)$ we can use the previous section to estimate all the $h(\rho_{m,p}^n)$ from the noisy linear combinations $y_n$. The algorithm can then be written as the following maximization-maximization scheme:
\begin{enumerate}
\item initialize $\beta_p = 1\quad\forall p=1,\dots,P$;
\item estimate $h$ at fixed $\beta$ by estimating GP conditional;
\item estimate $\beta$ at fixed $h$ using least squares;
\item optimize hyperparameters of GP kernel (if applicable)
\item stop if converged or go back to 2).
\end{enumerate}
\subsection{Hyperparameter optimization}
The chosen kernel may have hyperparameters that change its behaviour, which need to be adapted in order to best fit the HRF shape. If the covariance kernel $k$ depends on a hyperparameter $\vartheta$, then we optimize $k_\vartheta$ at every step we estimate $h$.

In a general framework, it is possible to set the hyperparameters by maximizing the marginal likelihood: 
$$\log p(y | \rho, \vartheta) = -\frac{1}{2}y^T {\Sigma_\vartheta^{11}}^{-1} y -\frac{1}{2}\log \text{det}(\Sigma_\vartheta^{11}) - \frac{n}{2}\log 2\pi,$$
The partial derivative of the marginal log-likelihood $\log p(y | X, \vartheta)$ with respect to each hyperparameter $\vartheta_j$ in e.g. $\vartheta = \{\gamma, C\}$ reads
$$\frac{\partial}{\partial \vartheta_j} \log p(y|\rho, \vartheta) = \frac{1}{2} \text{Trace}\left((\alpha \alpha^T - \Sigma_\vartheta^{-1})\frac{\partial \Sigma_\vartheta}{\partial \vartheta_j}\right),$$ 
with $\alpha = \Sigma_\vartheta^{-1} y$ the dual coefficients \cite{Rasmussen04}. Since our measurement kernel is a linear combination of HRF kernel evaluations, the derivatives in the parameters follow these same linear combinations.

For gradient descent, a useful early stopping criterion is increasing Leave One Out Error, which is less prone to overfitting than pure loglikelihood on same data.
\section{Experiments and Results}
In this work, a Gaussian kernel imposes smoothness:
$$k_\gamma(x, y) = C\exp\left(-\frac{1}{\gamma}\|y - x\|^2\right)$$
For the GP mean, one can use a certain HRF shape (e.g. gamma difference) or 0. At high noise levels the GP solution shrinks towards this mean function.
\subsection{Synthetic experimental data}
We generated synthetic data using an event-related paradigm (200 events) with 6 different event types, jittered presentation on average every 6s, TR$=2~$secs, and a gamma-difference HRF shape. 

\subsection{HRF recovery}

\begin{figure}
  \centering
  $\sigma_n=0.01$, SNR$=40~$dB\\
      \includegraphics[width=0.35\textwidth]{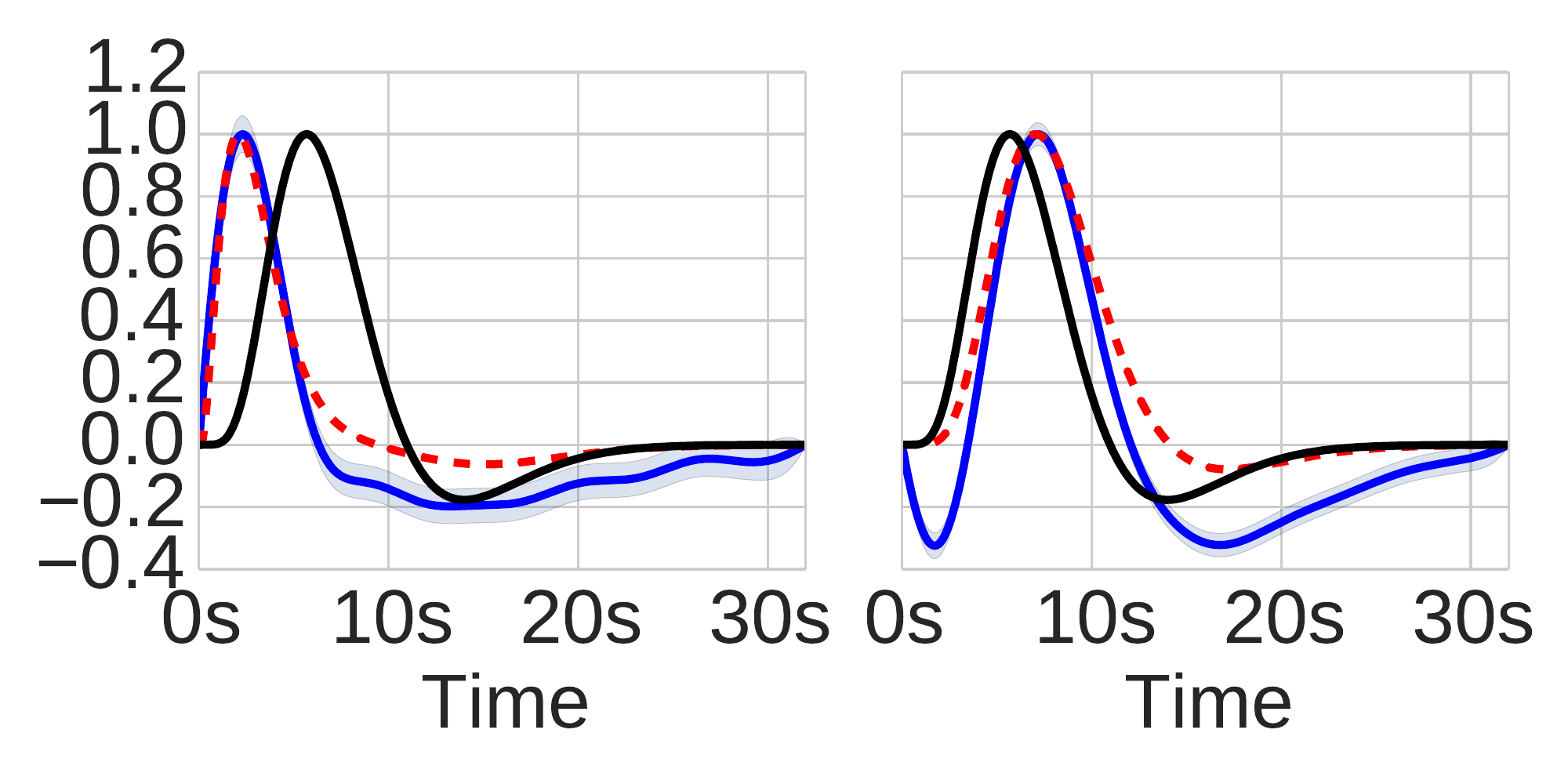}
    \\
  $\sigma_n=2$, SNR$=1~$dB \\
    \includegraphics[width=0.35\textwidth]{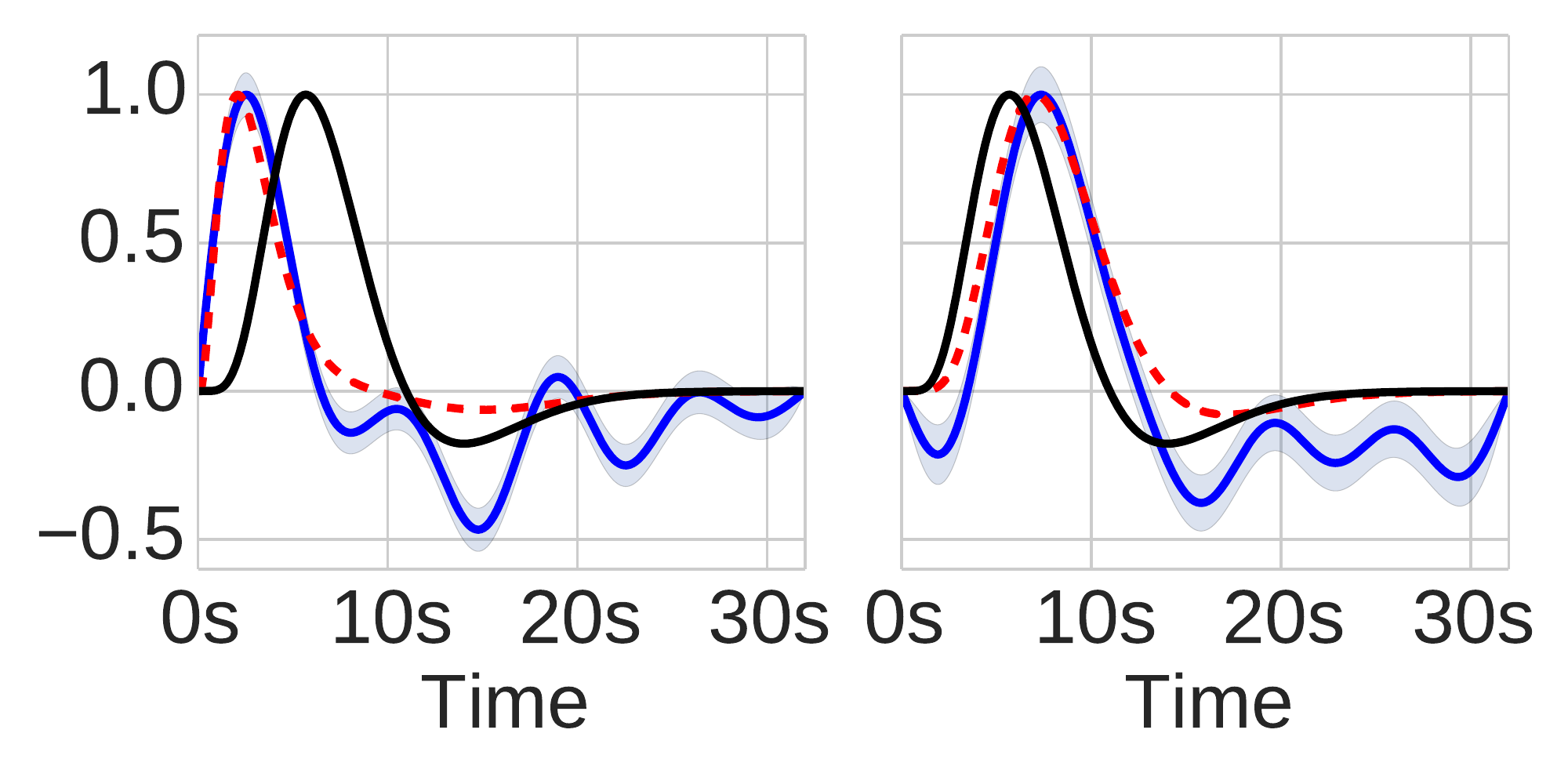}
    \\[-0.2cm]
    \caption{\textbf{Estimation of the HRF by GP on simulated data:} Dashed red line corresponds to the true HRF, black line to the GP mean function, and blue line to the estimated HRF. Using different noise levels: \emph{(top)} low noise (40 dB), and \emph{(bottom)} noisy signal (1 dB). \label{fig:hrfs}}
    \vspace{-20pt}
\end{figure}

The estimated HRF with GP method is shown in figure~\ref{fig:hrfs}, for scenarios with SNR 1 dB and 40 dB. We generate data with different peaks (3 and 8 seconds), and estimate them using a GP mean with peak at 5 seconds. The estimated HRFs are accurate in the peak, but incur noise in the tail at low SNR.

\subsection{Signal estimation accuracy}

\begin{figure}
  \centering
    \includegraphics[width=0.48\textwidth]{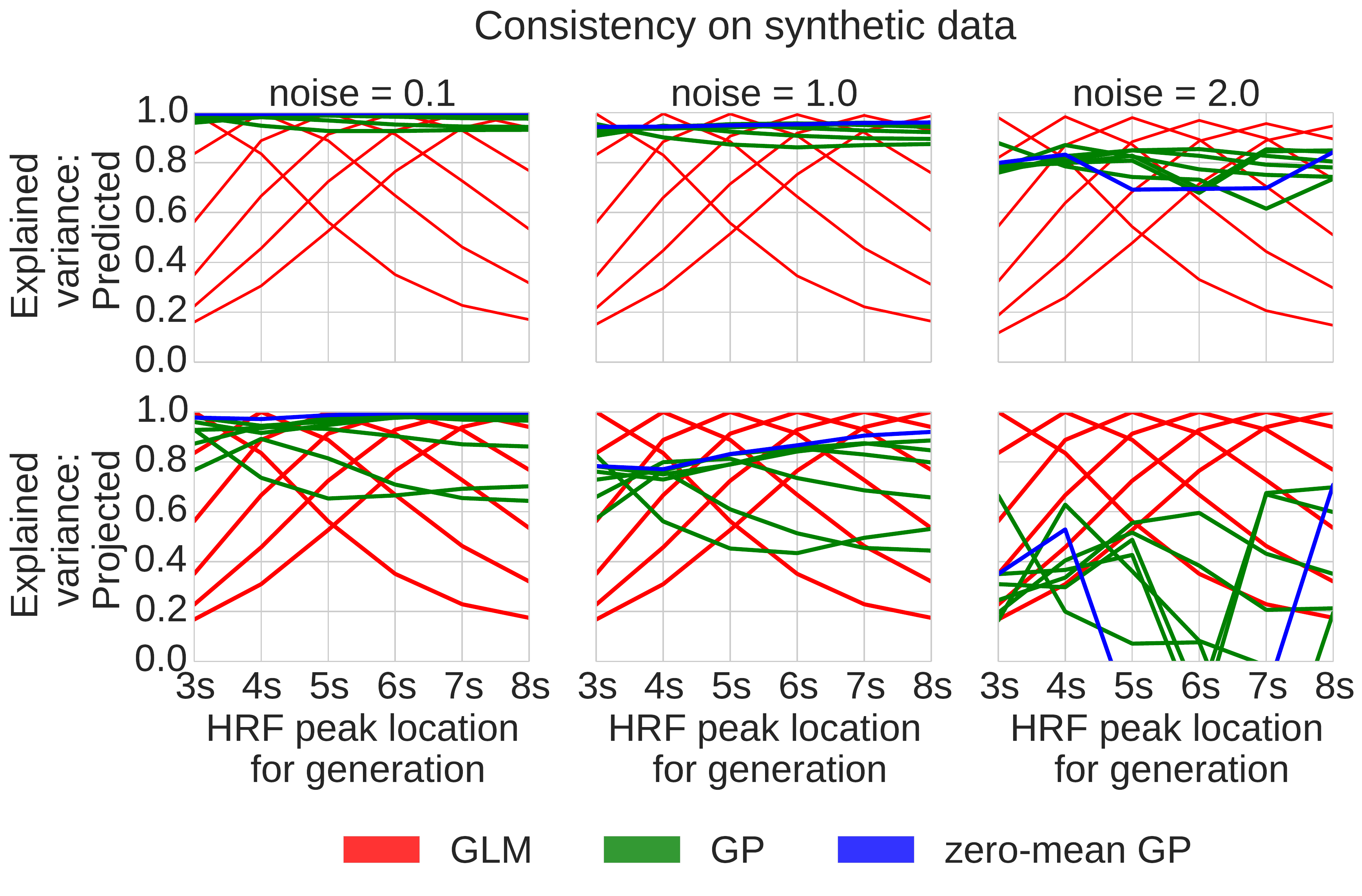}
    \caption{\textbf{Explained variance for different GLM estimators:} Each point on the x-axis represents a synthetic dataset with HRF peak at the indicated time. Each line represents a method, evaluated on each of these datasets. Red lines represent classic GLM with fixed HRF peaking at different points. Explained variance is at maximum when estimation and data-hrf correspond. Green lines show GP HRF estimations with mean HRF set to the same ones as used in the classic GLM. The blue line indicates GP HRF estimation using a zero-mean GP. \emph{(top)} shows explained variance on held-out data: The GLM method is used to estimate activation coefficients and HRF if applicable. The activation coefficients are used to predict on held-out data. 
\emph{(bottom)} Shows projection score: A generated fMRI timecourse is projected onto the span of its design matrix with varying HRF. This quantifies the goodness of the HRF independently from activation maps. The columns represent three SNR levels 0.1, 1.0 and 2.0. \label{fig:glmvsgp}}
\vspace{-15pt}
\end{figure}

Fig.\ref{fig:glmvsgp} evaluates the predictive capacity of classic GLM and GLM with estimated HRF. Several noisy synthetic datasets as described above are generated with gamma difference HRFs peaking at $t_p \in\{ 3, 4, \dots, 8\}$ seconds. The activations are estimated using several analysis methods: a) Classic GLMs using a set of design matrices with gamma difference HRFs peaking at all $t_p$; b) Gaussian process HRF estimation using $t_p$-peaking gamma-difference HRFs as mean functions; c) Gaussian process HRF estimation using a 0 mean function. 
Estimation error is quantified in two ways: 1) Predictive capacity on new data using estimated HRF and activation maps; 2) Projective capacity on new data: The estimated HRF is used to perform a GLM analysis on the held-out data and the residuals analyzed. Both analysis methods are performed at several noise levels for all the mentioned GLMs. In Fig. \ref{fig:glmvsgp}, each line corresponds to a method, each x-value to a dataset.

It is to be expected that when the classical GLM HRF peak location corresponds to that used for data generation, the error in both prediction and projection are very low, leading to perfect $R^2$ score. This changes as soon as there is discrepancy between the two HRFs. On the other hand, almost independently of the chosen mean HRF, the GP estimate of the HRF leads to better predictive and projective scores than the classic GLM. Zero-mean HRF performs similarly well.

\subsection{Impact of $\gamma$ parameter}
\begin{figure}[h]
\vspace{-10pt}
\centering
\includegraphics[width=0.48\textwidth]{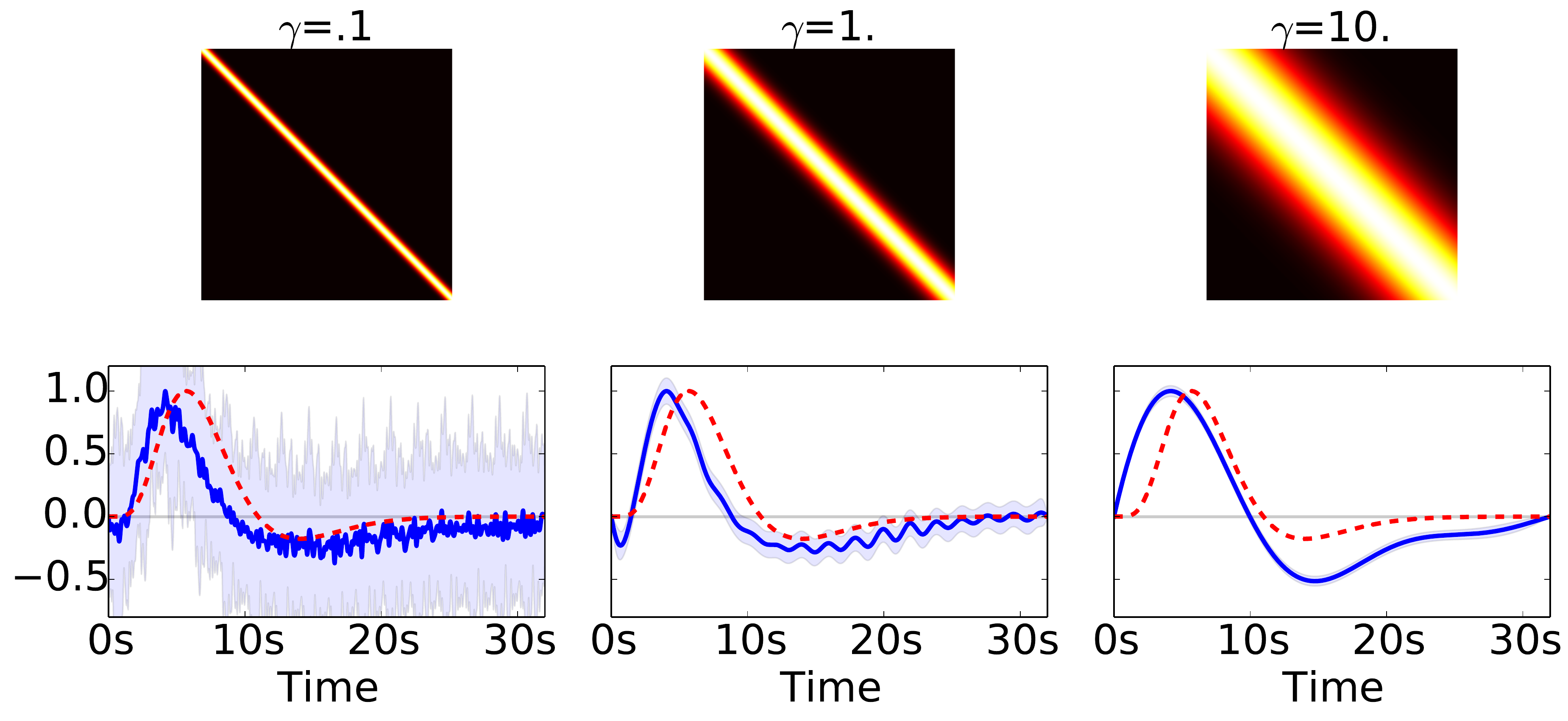} 
\caption{\textbf{Illustration of the impact of the $\gamma$ parameter on the estimation of the HRF:} In this experiment, the noise is fixed to $\sigma_n = 0.01$ (low-noise regime), without carrying out the parameter optimization. \emph{(top)} Shows the kernel matrix $k_\gamma$ for various parameters; \emph{(bottom)} shows the estimated HRF using that kernel. Dashed red line corresponds to the simulated HRF, and blue line to the estimated HRF. \label{fig:gamma}}
\end{figure}

The kernel $\gamma$ parameter encodes the smoothness of the estimated HRF. Fig.\ref{fig:gamma} shows the kernel matrices and the estimated HRFs corresponding to different values of $\gamma$. Higher values of $\gamma$ will enforce smoothness.

This parameter can be fixed according to prior knowledge or optimized by gradient ascent on the loglikelihood. 

\subsection{Tests on fMRI data}
We performed some experiments on the dataset AINSI~\footnote{http://thalie.ujf-grenoble.fr/ainsi}. The  experiment was designed
to map auditory, visual and motor brain function and consisted of $N = 142$ scans lasting $TR = 3$~s, with $TE = 33$~ms, FoV $220$~mm, each
yielding a 3-D volume. Two runs were acquired with different resolution of $2 \times 2 \times 4 $ mm$^3$ ($79 \times 95 \times 34$ voxels) and $3 \times 3 \times 3 $ mm$^3$ ($53 \times 63 \times 46$ voxels).  They were resliced to have the same resolution.
The paradigm was a fast event-related design (mean $ISI=5.1$~s) comprising sixty auditory
  and visual stimuli, mixed with computation, speech and motor tasks. Using a classic GLM with canonical HRF, voxels of the visual, auditory and motor regions with strong responses to the design were selected. The average over approximately 25 neighboring voxels around them was performed to decrease SNR of the timeseries.

\begin{figure}
  \centering
  \includegraphics[width=0.45\textwidth]{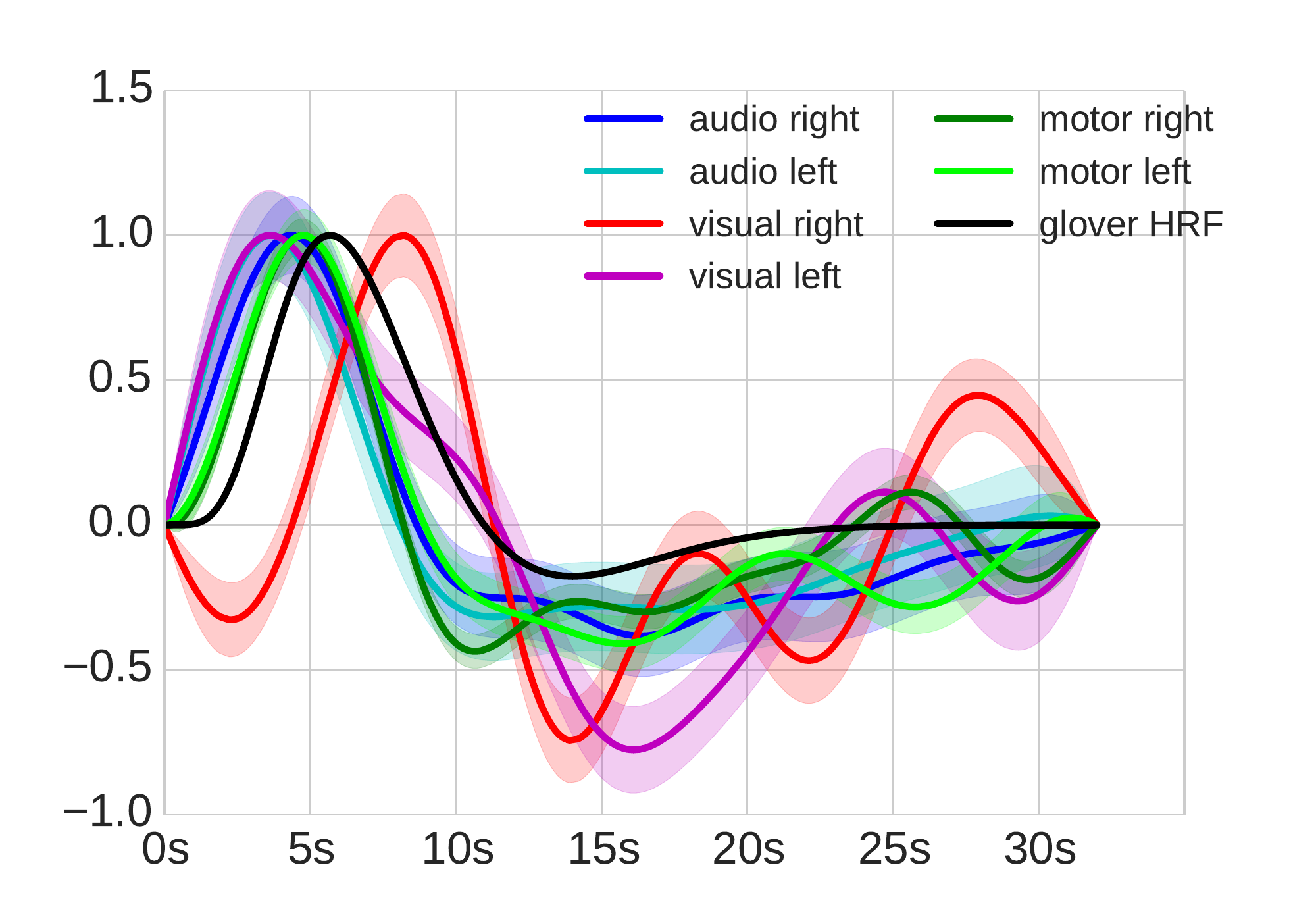} \\[-.2cm]
 \small{ Time }
    \caption{\textbf{Estimated HRF responses on 6 different regions of a single subject:} auditory, visual and motor cortices, left and right. The estimation was done on the mean timeseries over approximately 25 voxels around the voxel that responded strongest to the design matrix with canonical hrf in each region. In this case, $\gamma = 4$. \label{fig:hrf_regions} }
    \vspace{-20pt}
\end{figure}

First, we analysed the first BOLD run with the proposed GP approach. The estimated HRF for the different regions are depicted in Fig. \ref{fig:hrf_regions}. 

For quantitative validation we perform a prediction on the second BOLD run for the same subject using the estimated activations and HRF on the paradigm of the second run. To compare, we also fit a classic GLM with canonical HRF on the first run and use the activations to predict on the second. We call this setting \textit{prediction}.
In order to evaluate the HRF in an isolated manner,  we also run a GLM using the estimated HRF found with GP ($h_{GP}$) on the second run and evaluate the in-sample prediction on the second run compared to doing the same with a canonical HRF ($h_{can}$). We call this setting \textit{projection}, because we quantify the capacity of the column span of the matrix to model the signal.

In both cases, comparison measure between predicted timecourse and true timecourse is Pearson correlation. Prediction results can be found in table~\ref{table:prediction}. In \textit{prediction}, a classical GLM performs better than GP. In \textit{projection} the HRF estimated with GP we can improve the prediction in most of the voxels.
\begin{table}
\caption{Prediction results on held-out data for 6 timeseries corresponding to 6 different regions of interest. \label{table:prediction}}
\vspace{-.2cm}
\begin{center}
\begin{tabular}{| p{1.7cm} | C{1.2cm} C{1.2cm} |C{1.2cm} C{1.2cm} |}
\hline
 & \multicolumn{2}{|c|}{Prediction} & \multicolumn{2}{|c|}{Projection}\\ \cline{2-5}
\begin{tabular}{l}
ROI \end{tabular}
& \begin{tabular}{c}
 $h_{can}$ \end{tabular}
& GP & 
\begin{tabular}{c}
$h_{can}$ \end{tabular}
& \begin{tabular}{c} 
with $h_{GP}$ \end{tabular}
\\ \hline
Auditory right & 0.51 & 0.41 & 0.66 & 0.56  \\ 
Auditory left & 0.48 & 0.56 & 0.65 & 0.70 \\ 
Visual right & 0.10 & 0.12 & 0.39 & 0.44 \\ 
Visual left & 0.32 & 0.30 & 0.44 & 0.46 \\ 
Motor right & 0.59 & 0.51 & 0.68 & 0.58 \\ 
Motor left & 0.57 & 0.63 & 0.69 & 0.72  \\
\hline
\end{tabular}
\end{center}
\vspace{-20pt}
\end{table}
\vspace{-10pt}
\section{Discussion}
We have proposed a non-parametric method that models the HRF as a continuous function using a Gaussian Process. 

As it has been shown in figure~\ref{fig:glmvsgp}, the standard approach for analysing fMRI data, the GLM with fixed HRF, is less accurate when the true HRF is different from the one used to estimate the activations. Since it is well-known that the HRF varies for different regions and subjects, a detailed analysis can benefit from taking this into account. 

This work is a proof of concept. Results on simulated data suggest that the approach is well defined and it is an alternative for HRF estimation. However, results on real data suggest the need for further investigation. High noise level and the tendency of GP to overfit are an issue. Note however that our selection procedure favors voxels well modeled by classic GLM. Pooling HRF over several voxels may help in general~\cite{Makni05}.

Several directions of future work become apparent. Detailed validation and comparison of the method to other candidate methods is necessary. The investigation of other kernels should be fruitful, since length-scale is not uniform across time. The maximization-maximization-approach can be replaced by variational expectation-maximization leading to more robust convergence.





\bibliographystyle{IEEEtran}
\bibliography{myBib}

\begin{thebibliography}{10}
\providecommand{\url}[1]{#1}
\csname url@samestyle\endcsname
\providecommand{\newblock}{\relax}
\providecommand{\bibinfo}[2]{#2}
\providecommand{\BIBentrySTDinterwordspacing}{\spaceskip=0pt\relax}
\providecommand{\BIBentryALTinterwordstretchfactor}{4}
\providecommand{\BIBentryALTinterwordspacing}{\spaceskip=\fontdimen2\font plus
\BIBentryALTinterwordstretchfactor\fontdimen3\font minus
  \fontdimen4\font\relax}
\providecommand{\BIBforeignlanguage}[2]{{%
\expandafter\ifx\csname l@#1\endcsname\relax
\typeout{** WARNING: IEEEtran.bst: No hyphenation pattern has been}%
\typeout{** loaded for the language `#1'. Using the pattern for}%
\typeout{** the default language instead.}%
\else
\language=\csname l@#1\endcsname
\fi
#2}}
\providecommand{\BIBdecl}{\relax}
\BIBdecl

\bibitem{Handwerker04}
D.~A. Handwerker, J.~M. Ollinger, and M.~D'Esposito, ``Variation of bold
  hemodynamic responses across subjects and brain regions and their effects on
  statistical analyses,'' \emph{Neuroimage}, vol.~21, 2004.

\bibitem{Badillo13}
S.~Badillo, T.~Vincent, and P.~Ciuciu, ``Group-level impacts of within- and
  between-subject hemodynamic variability in {fMRI},'' \emph{Neuroimage},
  vol.~82, 2013.

\bibitem{Pedregosa15}
F.~Pedregosa, M.~Eickenberg, P.~Ciuciu, B.~Thirion, and A.~Gramfort,
  ``Data-driven hrf estimation for encoding and decoding models,''
  \emph{NeuroImage}, vol. 104, 2015.

\bibitem{Goutte00}
C.~Goutte, F.~{\AA}. Nielsen, and L.~K. Hansen, ``Modeling the hemodynamic
  response in fmri using smooth fir filters,'' \emph{IEEE Trans. on Medical
  Imaging}, vol.~19, 2000.

\bibitem{Woolrich04}
M.~W. Woolrich, T.~E. Behrens, and S.~M. Smith, ``Constrained linear basis sets
  for hrf modelling using variational bayes,'' \emph{NeuroImage}, vol.~21,
  2004.

\bibitem{Kay08}
K.~N. Kay, S.~V. David, R.~J. Prenger, K.~A. Hansen, and J.~L. Gallant,
  ``Modeling low-frequency fluctuation and hemodynamic response timecourse in
  event-related fmri,'' \emph{Human brain mapping}, vol.~29, 2008.

\bibitem{Zhang14}
T.~Zhang, F.~Li, M.~Z. Gonzalez, E.~L. Maresh, and J.~A. Coan, ``A
  semi-parametric nonlinear model for event-related fmri,'' \emph{NeuroImage},
  vol.~97, 2014.

\bibitem{Makni05}
S.~Makni, P.~Ciuciu, J.~Idier, and J.-B. Poline, ``Joint detection-estimation
  of brain activity in functional mri: a multichannel deconvolution solution,''
  \emph{IEEE Trans on Signal Processing}, vol.~53, 2005.

\bibitem{Vincent10}
T.~Vincent, L.~Risser, and P.~Ciuciu, ``Spatially adaptive mixture modeling for
  analysis of {within-subject fMRI} time series,'' \emph{{IEEE Trans. on
  Medical Imaging}}, vol.~29, 2010.

\bibitem{Ciuciu03}
P.~Ciuciu, J.-B. Poline, G.~Marrelec, J.~Idier, C.~Pallier, and H.~Benali,
  ``Unsupervised robust nonparametric estimation of the hemodynamic response
  function for any fmri experiment,'' \emph{IEEE Trans. on Medical Imaging},
  vol.~22, 2003.

\bibitem{Rasmussen04}
C.~E. Rasmussen, ``Gaussian processes in machine learning,'' in \emph{Advanced
  lectures on machine learning}, 2004.

\end{thebibliography}

\end{document}